\documentstyle{elsart} 

\input epsf

\begin{document}

\begin{frontmatter}
\title{Current algebra and soft pionic modes in asymmetric quark matter} 
\thanks{Research supported by: PRAXIS XXI/BCC/4299/94, 
        the Polish State Committee for
        Scientific Research 2P03B-188-09, PCERN/FIS/1034/95, and
        PESO/S/PRO/1057/95}
\author{Wojciech Broniowski}
\address{H. Niewodnicza\'nski Institute of Nuclear Physics,
         PL-31342 Krak\'ow, Poland}
\author{Brigitte Hiller}
\address{Centro de {F\'{\i}sica} Te\'orica, 
         Department of Physics, University of Coimbra, P-3000 
         Coimbra, Portugal}

\begin{abstract}
Pionic excitations are analyzed in isospin-asymmetric Fermi gas 
of constituent quarks. It is found that in addition to the usual pionic 
excitations, there exists a very low 
collective mode with quantum numbers of the 
charged pion (spin-isospin sound). 
In the chiral limit the excitation energy of one of the modes scales as the
current quark mass, and the mode saturates current-algebraic sum rules. 
This is in agreement with predictions of current algebra for 
isospin-asymmetric medium.    
\end{abstract}

\end{frontmatter}

Recently chiral current algebra has been applied 
\cite{CB,Lutz} to systems with finite chemical potentials 
in order to study excitations in such systems. 
A formal prediction made by Cohen and one of us (WB) is that in the
isospin-asymmetric nuclear medium there exists an excitation with
quantum numbers of the charged pion ($\pi^+$ for medium with negative
isospin density), which in the 
chiral limit becomes very soft \cite{CB}. 
Its excitation energy is proportional to
the current quark mass $m$, and not $\sqrt{m}$ as in the case
of the vacuum, and in this sense the mode 
is softer than the usual pseudo-Goldstone boson.\footnote{
Note that in isospin-asymmetric medium the Goldstone theorem 
does not apply for charged excitations, since the constraint breaks explicitly 
the appropriate symmetry \cite{CB}.}  
If such a soft mode exists at the physical value of $m$ in dense nuclear
medium, and if it is strongly coupled, it would bring 
important phenomenological
consequences by dominating excitation functions and thermal properties 
of the system at low energies. Unfortunately, current algebra gives us no hint
as to whether the physical value of $m$ is low 
enough such that the formal results
of Ref.~\cite{CB} have practical importance. 
After all, in a nuclear system
we have additional energy scales, {\em e.g.} chemical potentials or energies 
of single-particle excitations, and $m$, or rather the vacuum value
of the pion mass, $m_\pi$, need not be small compared to these scales. 

In order to gain some quantitative insight we study a model
in which the asymmetric nuclear medium is composed 
of the Fermi gas of constituent
quarks of the Nambu--Jona-Lasinio model \cite{NJLrev}. 
Our results are as follows: 
Firstly, since the model obeys all requirements of 
current algebra, the chiral soft
mode emerges when $m$ is decreased. However, this mode is not 
necessarily a continuation of the vacuum pion branch. 
Another branch of low excitations exists ($\sim 10$~MeV), 
which is analogous to the
{\em spin-isospin sound} \cite{migdal,rhowil} found in conventional 
approaches to neutron matter.
We show that this mode 
contributes sizably to current-algebraic sum rules, also at the physical
value of $m$.
  
Let us begin with a reminder of the sum rules of Ref.~\cite{CB,Lutz}:
\begin{eqnarray}
\!\!\!\!\!\!\!\!\!\!\!\! -2 m \langle \bar{q} q \rangle_C & = &
\sum_{j^-} {\rm sgn}(E_{j^-}) 
\left | \langle j^-\!\! \mid \! J^-_{5,0}(0) \!\!
\mid \!\! C \rangle \right |^2 
+  \sum_{j^+} {\rm sgn}(E_{j^+}) 
\left | \langle j^+\!\! \mid \! J^+_{5,0}(0) \!\! 
\mid \!\! C \rangle \right |^2 ,
\label{eq:GMOR} \\
2 \rho_{I=1} & = & 
\sum_{j^-} \frac{1}{\mid \!\! E_{j^-}\!\! \mid} 
\left | \langle j^-\!\! \mid \! J^-_{5,0}(0) \!\! 
\mid \!\! C \rangle \right |^2 
- \sum_{j^+} \frac{1}{\mid \!\! E_{j^+}\!\! \mid} 
\left | \langle j^+\!\! \mid \! J^+_{5,0}(0) \!\! 
\mid \!\! C  \rangle \right |^2 \;,
\label{eq:iso}
\end{eqnarray}
where
$J^a_{5,0} = \bar q \gamma_0 \gamma_5 \half \tau^a q$ is the time component 
of the axial vector current, $\mid \!\! C \rangle$ denotes 
a uniform medium subject to constraints
imposed externally, $\mid j^\pm \rangle$ are 
excited states {\em with three-momentum equal zero} 
of isospin $\pm 1$, and
$E_{j^\pm}$ are their excitation energies.
The constraints are the baryon density, $\rho_B$, 
and the  isospin
density $\rho_{I=1}$. 
In Eq.~(\ref{eq:GMOR}) we recognize the generalization 
of the Gell-Mann--Oakes--Renner
sum rule for finite densities.
The ``isovector'' sum rule (\ref{eq:iso}) is nontrivial 
only for systems with isospin asymmetry.
As shown in Ref.~\cite{CB}, since the isovector 
chemical potential remains finite as $m \to 0$,
in a medium with negative isospin density 
only positive-isospin modes can contribute  
to the above sum rules in the chiral limit.
This implies, under some additional weak
assumptions, the existence of a ``chiral soft mode'' $\mid + \rangle$
 for which
\begin{equation}
E_+ \sim m ,\;\;\;\;\;\;\;\; 
\left | \langle + \mid J_{5,0}^+ \mid C \rangle \right | \sim \sqrt{m} 
\label{eq:soft}
\end{equation}
This mode saturates both sum rules in the chiral limit.
On the other hand, there are no soft modes with negative isospin, 
and $E_{j^-} \sim 1$. 
The presented  behavior is radically different from the case of the vacuum or
symmetric matter, where the lowest $\pi^+$ and $\pi^-$ excitations
scale as $\sqrt{m}$. The $\pi_0$ excitations scale in all cases 
as $\sqrt{m}$ \cite{CB}. 

Now we pass to an illustration of these general results in a simple model: the 
Nambu--Jona-Lasinio model with scalar and vector interactions 
\cite{NJLrev,BernardAP}.
The model is consistent with current algebra \cite{Lutz}, and 
its variants have been applied extensively to describe both meson and
baryon physics. The Lagrangian is
\begin{eqnarray}
\!\!\!\!\!\!\!\!\! {\cal L} = 
 \bar q({\rm i} \not \! \partial - m)q &+& 
  {\frac{G_\sigma}{2}}\left( (\bar q q)^2 + 
   (\bar q{\rm i}\gamma_5 \tau^a q)^2\right) 
+ {\frac{G_\delta}{2}}\left( (\bar q \tau^a q)^2 + 
   (\bar q{\rm i}\gamma_5 q)^2\right) \nonumber \\
&-& {\frac{G_\rho}{2}}\left( (\bar q \gamma_\mu \tau^a q)^2 + 
   (\bar q \gamma_5 \gamma_\mu \tau^a q)^2\right) 
- {\frac{G_\omega}{2}} (\bar q \gamma_\mu q)^2  \;.   
\label{eq:lag}
\end{eqnarray}
%
Using the Hartree approximation one arrives at 
self-consistency equations for the values of the
scalar-isoscalar field $S$, the scalar-isovector field $\delta$, 
the time component of the neutral vector-isovector field $\rho$,
and the time component of the vector-isoscalar field $\omega$ :
%
\begin{eqnarray}
S &=& m - G_\sigma \langle \overline{u} u + \bar d d \rangle, \; 
\delta = - G_\delta \langle \overline{u} u - \bar d d \rangle, \nonumber \\ 
\rho &=& 2 G_\rho \langle u^+ u - d^+ d \rangle, \;
\omega =  G_\omega \langle u^+ u + d^+ d \rangle.
\label{eq:fields}
\end{eqnarray}
The scalar and vector densities of the $u$ quark are equal to 
(analogously for the $d$ quark)
\begin{equation}
\!\!\!\!\!\!\!\!
\langle \overline{u} u \rangle = 2 N_c \int \frac{d^3k}{(2\pi)^3} 
\frac{M_u}{\sqrt{k^2+M_u^2}} \Theta_u, \;\;\;\;
\langle u^+ u \rangle = 2 N_c \int \frac{d^3k}{(2\pi)^3} \Theta(k_u-|k|) \;.
\label{eq:dens}
\end{equation}
We have defined 
$\Theta_u=\Theta(k_u-|k|)-\Theta(\Lambda-|k|)$,
where $\Lambda$ is the 
sharp three-momentum cut-off used in this paper,
and $k_u$ and $k_d$ are
the $u$ and $d$ quark Fermi momenta. 
We have also introduced scalar self-energies   
$M_u=S+\delta$ and $M_d=S-\delta$. Self-consistency requires that
the quark propagators are evaluated with mean-fields (\ref{eq:fields}):
\begin{equation}
S_{u/d}^{-1} = \not \! p - \gamma_0 (\pm \frac{\rho}{2} + \omega) - M_{u/d} 
 + i \varepsilon \, {\rm sgn}(\mu_{u/d}-p_0) \;, 
\label{eq:S}
\end{equation}
where $\mu_u$ and $\mu_d$ are the chemical potentials 
of the $u$ and $d$ quarks.
Next, we consider the charged pion propagator.
We choose $k_d > k_u$, as for instance in neutron matter.
Were there no vector-isovector 
meson effects ({\em i.e.} $G_\rho = 0$), 
the inverse pion propagator
would have a simple form $1-G_\sigma J_{PP}$, where  
$J_{PP} = -i {\rm Tr} \int \frac{d^4k}{(2\pi)^4} 
 \gamma_5 S_u(p+\half q) \gamma_5 S_d(p-\half q)$.
However, in presence of vector-isovector coupling 
there occurs $\pi-A_1$ mixing, and to find excitation energies 
with pion quantum numbers one has to find
{\em zeros} of the determinant of the inverse $\pi-A_1$ propagator matrix
(see {\em e.g.} Ref.~\cite{Lutz} for details).
In the considered model this determinant can be written as 
%
\begin{equation}
\!\!\!\!\!\!\!\!\!\! D(q_0) = \frac{q_0}{q_0-\rho}
\left \{\frac{m}{S} + \left ( \frac{4 m S G_\sigma^{-1} G_\rho }{q_0(q_0-\rho)}
- 1 \right) \left [
G_\sigma J_{PP}(q_0)- \frac{S-m}{S} \right ] \right \} \;,
\label{eq:det}
\end{equation}
where $q_0$ is the frequency of the excitation, and
\begin{eqnarray}
J_{PP}(q_0) &=& 4 N_c \int_{k_u}^\Lambda
\frac{d^3k}{(2\pi)^3}
\frac{(\rho-q_0)+ 2\delta M_u/\sqrt{k^2+M_u^2}}
{(\rho-q_0)^2 + 2 (\rho-q_0)\sqrt{k^2+M_u^2} + 4 S \delta} \nonumber \\ 
&+&(u \to d,\;\; \rho \to -\rho,\;\; \delta \to - \delta,\;\; q_0 \to - q_0) \;.
\label{eq:jpp}
\end{eqnarray}
%
Let us look at the analytic
structure of $D$, or $J_{PP}$, in $q_0$. 
The state $\mid \!\! C \rangle$ 
consists of the Fermi seas of $d$ and $u$ quarks, with
$k_d > k_u$, as well as of
the Dirac sea occupied down to the cut-off $\Lambda$. 
A positive-charge Fermi sea excitation 
moves a quark from the occupied 
$d$ level to an unoccupied $u$ level. Pauli blocking allows this 
when \mbox{$\rho + \sqrt{k_d^2 + M_u^2} - \sqrt{k_d^2 + M_d^2}
< q_0 < \rho + \sqrt{k_u^2 + M_u^2} - \sqrt{k_u^2 + M_d^2}$}.\\ 
Thus, within these
boundaries $D(q_0)$ possesses a cut. 
There are also additional cuts 
corresponding to quark-antiquark creation, but their effects are negligible
in our study.
As explained {\em e.g.} in \cite{migdal,rhowil} in the framework of 
conventional nuclear physics, it is possible for 
the pion propagator in neutron matter
to have an additional pole at very low excitation energies:
the {\em spin-isospin sound}. We will show that an analogous
phenomenon occurs in our model.   

For the numerical study we use the following parameter sets: 
\begin{eqnarray}
\!\!\!\!\!\!\!\!\!\!\!\!\!\!\!\!\!\!\!\!\!\!\! {\rm I:}&&
G_\sigma = 7.55{\rm GeV}^{-2},\;  G_\delta = 5.41{\rm GeV}^{-2},\;
G_\rho = 7.09{\rm GeV}^{-2},\;\nonumber \\ 
&& \Lambda = 750{\rm MeV},\; m = 3.52{\rm MeV}, \nonumber \\
\!\!\!\!\!\!\!\!\!\!\!\!\!\!\!\!\!\!\!\!\!\!\! {\rm II:}&& 
G_\sigma = 4.35{\rm GeV}^{-2},\;  
G_\delta = 3.34{\rm GeV}^{-2},\;
G_\rho = 12.4{\rm GeV}^{-2},\; \nonumber \\ 
&& \Lambda = 954{\rm MeV},\; m = 2.03{\rm MeV}.\nonumber
\label{eq:par}
\end{eqnarray}
Set I is the SU(3) fit of Ref.~\cite{Klimt}, which fits the 
$m_\pi = 139$MeV, $F_\pi=93$MeV, $m_\eta=519$MeV, and $m_\rho = 765$MeV.
This leaves one parameter out of the original 
five undetermined, which is expressed
through the constituent quark mass in the vacuum, set
arbitrarily to $M = 361$MeV.
Set II also fits $m_\pi$, $F_\pi$, $m_\eta$, and $M = 361$MeV, but not 
$m_\rho$. Instead, we chose
a much larger value of $G_\rho$. Such greater values are needed if,
for instance, 
one wishes to fit the $a_{11}$ pion-pion scattering length \cite{BernardAP}.
Since the excitations carry no baryon number, 
the value of $G_\omega$ is irrelevant.

We define $x=\rho_B/\rho_0$, where
$\rho_B$ is the baryon density and $\rho_0 = 0.17 {\rm fm}^{-3}$ is the 
nuclear saturation density. The relative concentration of $d$ quarks
is denoted by $y = \rho_d/(\rho_u+\rho_d)$.
For the results presented below we set $y=2/3$, which
corresponds to pure neutron matter. With $y$ fixed, we vary the $x$ variable, 
increasing the density from 0 to a few times the saturation density.
Figure 1 shows the results for the parameter set I.
In (a) we plot the mean fields (\ref{eq:fields}).
In (b) we show the excitation energies of
what we call the ``big'' modes, the ones that at $x=0$ (vacuum) 
become the usual
pions. The excitation energy of $\pi^+$ drops to about 100MeV 
at $x=2$, and then
increases again
at higher $x$. The excitation energy of $\pi^-$ increases 
sharply, which is a result of Pauli blocking.
At low densities the behavior of these excitation energies is
controlled by the Weinberg-Tomosawa term in the $\pi-N$ 
scattering \cite{Lutz}. The lower part of
Fig. 1(b) shows the excitation energy of the {\em third}
solution of the equation $D(q_0)=0$. This mode
is the {\em spin-isospin sound}, $\pi^+_s$.  
Its energy is small, rising up
to only 6MeV at $x=2$. The shaded area shows the 
cut discussed after Eq.~(\ref{eq:jpp}).
The  spin-isospin sound emerges from the cut at 
a very low but finite value of $x$. 
A similar phenomenon has also been obtained in studies of the kaon propagator
in nuclear matter \cite{sao,kubodera}. 
How relevant is this mode? 
A natural measure of its strength is provided by the contributions 
to the sum rules (\ref{eq:GMOR}-\ref{eq:iso}).
Explicitly, we have
\begin{equation}
\!\!\!\!\!\! {\rm sgn}(E_{j^\pm}) 
\left | \langle j^\pm\!\! \mid J^\pm_{5,0} \mid C \rangle \right |^2 =
- \left . \frac{2m}{q_0-\rho} 
\frac{\left [ S J_{PP}(q_0)-G_\sigma^{-1} (S-m) \right ]}
{dD(q_0)/dq_0} \right |_{q_0=E_j^\pm} \;.
\label{eq:expole}
\end{equation}
The contributions of the three modes to the sum rules 
are plotted in Fig.~1(c-d).  
Whereas in the GMOR sum rule the spin-isospin sound 
contributes up to 10\% 
(which is still noticeable), 
in the isovector sum rule it dominates over 
the ``big'' modes at $x > 2$, and at $x=3$ practically 
saturates the sum rule. In that sense the mode is strong.  
Note that the three modes satisfy the sum 
rules at the 99\% level, which means
that the contributions from cuts are tiny. 
Next, we present a formal study of the chiral limit.
In Fig.~2(a) we plot the excitation energies of the three modes
at $x=2$ and $y=2/3$ as functions of $\alpha$, which is the ratio of the 
current quark mass to its physical value from the parameter set I. 
Hence $\alpha=1$ corresponds to the physical case, and $\alpha=0$ is the
strict chiral limit. 
The spin-isospin sound becomes the chiral  
soft mode of Eq.~(\ref{eq:soft}), but only at $\alpha < 10^{-4}$ (!), where 
its excitation energy scales linearly with $m$. 
For higher values of $\alpha$ the energy remains flat. The next two
plots, Fig.~2(b-c), show the contributions to the sum rules
as a function of $\log \alpha$.
As required by Eq.~(\ref{eq:soft}), the chiral soft mode saturates the sum 
rules in the chiral limit, and becomes dominant at $\alpha < 10^{-3}$.

Figure 3 shows the results for the parameter set II.
The important qualitative difference when compared to 
Fig.~1  may be seen in the bottom of Fig.~(c):
the excitation energy of the spin-isospin sound is negative now. 
As discussed in Ref.~\cite{CB}, the energy of a charged excitation 
in asymmetric medium need not be positive. This is because we require that 
the state $\mid \!\! C \rangle$ be the ground state for a 
given value of the constraint,
here the isospin. A charged excitation connects 
$\mid \!\! C \rangle$ to states with
different isospin, hence such an excitation may in fact be a deexcitations.
More explicitly,   
we note that the $u$ quark is lighter than $d$, because the 
difference of their vector self energies, $\rho$, which is negative, 
overcomes the positive difference of the scalar self energies 
(cf. Eq~(\ref{eq:S})). Therefore changing
$u$ to $d$ is favorable energetically. 
Note that this negative excitation is still above the chemical potential 
for isospin, which is equal to 
\mbox{$\mu_u - \mu_d = \rho + \sqrt{k_u^2+M_u^2} - \sqrt{k_d^2+M_d^2}$}, 
hence it carries {\em positive} isospin. 
Thus the system may lower its energy by spontaneously
emitting the $\pi^+_s$ mode. However, by charge conservation 
this has to be accompanied by the emission of a particle of negative
charge. The strong-interaction modes ($\pi^-$, or ${\bar u} d$ pair) 
are heavy enough, so the system remains stable with respect to strong
decays. In weak decays the accompanying particle is the 
electron or muon. Depending on the value of the 
chemical potential of these particles, the reaction may lead 
to a net energy gain. In that case the system is unstable with respect
to weak interactions.

Figures 3(c-d) show the sum rules for the parameter set II. 
Again, we notice that the spin-isospin sound
mode contributes at the level of 10-20\% to the GMOR sum rule, but with
opposite sign. It dominates the isovector sum rule at $x>3$. 
Figure 4 shows a qualitatively different behavior
of the system close to the chiral limit  
for the parameter set II than for the set I. 
We see that now it is the ``big'' $\pi^+$ mode, not the $\pi^+_s$, 
which assumes the role
of the chiral soft mode. The $\pi^+_s$ has a finite energy in the
chiral limit, and does not contribute to the sum rules. These are dominated
by the ``big'' $\pi^+$ mode at $\alpha < 10^{-3}$, where it scales linearly 
with $m$.

The linear dependence of the excitation energy of the chiral soft mode 
on the current quark mass takes effect only at extremely small
values of $m$, corresponding to $m_\pi$ 10---100 times
lighter than the physical value. Hence the satisfaction of the 
laws (\ref{eq:soft}) has formal rather than practical meaning. Nevertheless, 
it is interesting that these laws may be fulfilled by the highly collective
spin-isospin sound, rather than by the vacuum pion branch. In our
model at all values of $m$ and densities the three pionic modes saturate 
the current-algebra sum rules at the 
99\% level, and the contributions from the spin-isopin sound are big.
This shows that in constructing an effective theory of mesons in medium
one should account for such excitations, including 
possible colective modes, and that
the contribution from particle-hole excitations are negligible. 
It is challenging to see if 
similar results hold in more realistic approaches to nuclear matter, 
with nucleon degrees of freedom. In particular, the role of
short-range correlations should be examined, since they are known affect 
strongly the excitation spectra in nuclear matter. 

We are grateful to Drs. A. Blin, 
J. da Providencia, M. C. Ruivo and C. A. de Sousa
for useful discussions. WB thanks the Centro de {F\'{\i}sica} Te\'orica
of the University of Coimbra for its hospitality.

\newpage

\pagestyle{empty}

\thispagestyle{empty}
\begin{figure}
\vspace{-12mm}
\epsfxsize = 9 cm
\centerline{\epsfbox{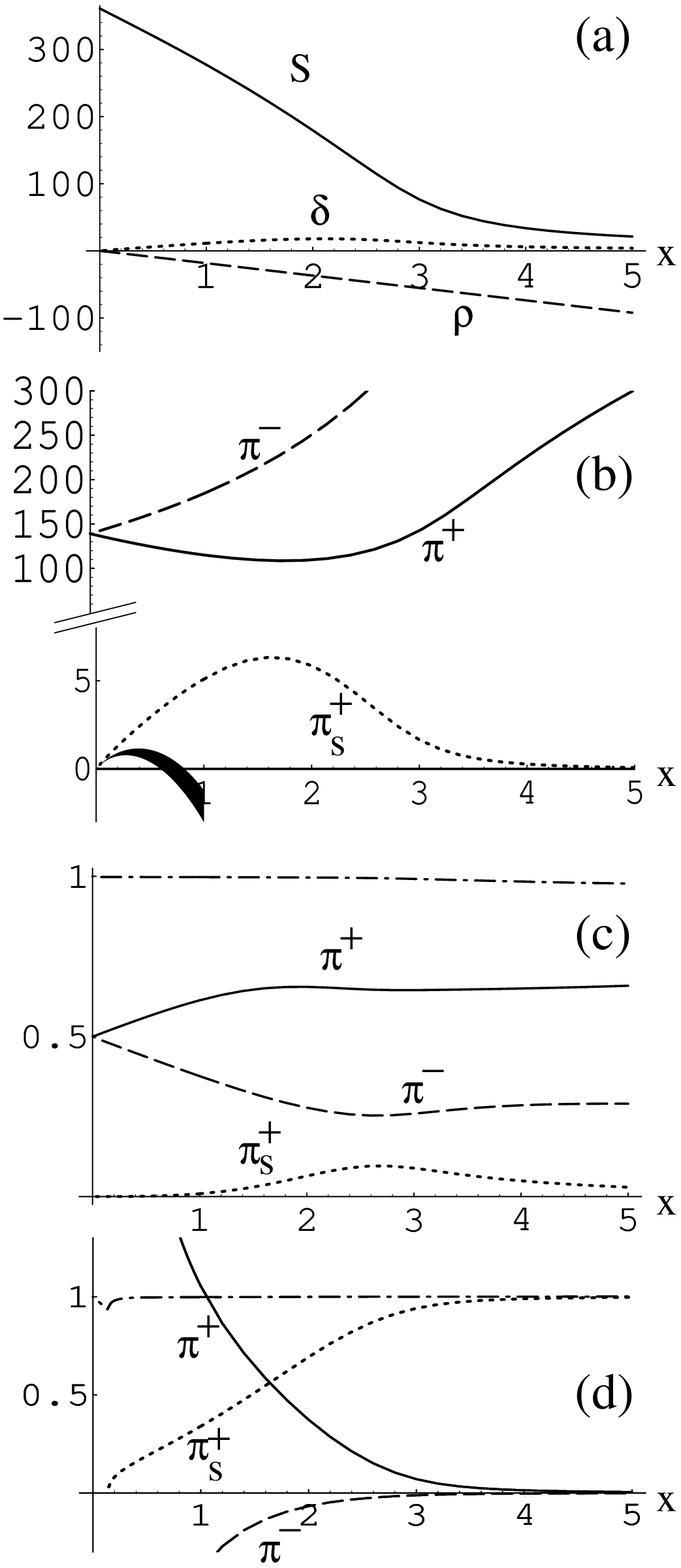}}
\vspace{-10mm}
\label{fig:setI}
\caption{ 
Dependence of various quantities on $x=\rho_B/\rho_0$
for the parameter set I and $y=2/3$:
(a) Mean fields in MeV.
(b) Excitation energies for the ``big'' modes (top) and
    the spin-isospin sound (bottom) in MeV. Also shown is the 
particle-hole production cut (shaded region).
(c) The relative  contributions of the three modes to the GMOR sum rule.
The sum (dash-dot line) is indistinguishable from one.
(d) Same as (c) for the isovector sum rule.
}
\end{figure}

\newpage

\thispagestyle{empty}

\begin{figure}
\vspace{-12mm}
\epsfxsize = 9 cm
\centerline{\epsfbox{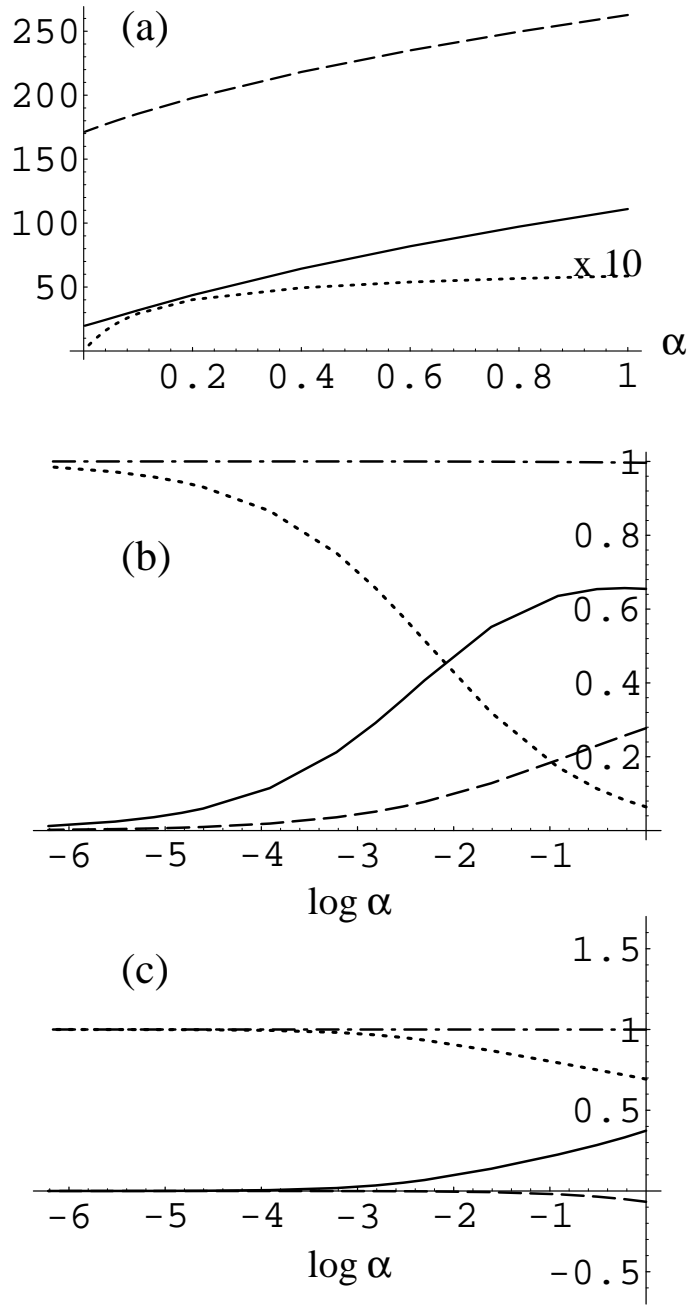}}
\vspace{-10mm}
\label{fig:chirI}
\caption{Study of the chiral limit. 
Convention for lines as in Fig.~1(b). 
(a) Dependence of excitation energies of the three modes
on the ratio of the current quark mass to its physical value, $\alpha$.
The curve for the spin-isospin sound has been multiplied by 10.
(b) Relative  contributions of the three modes to the GMOR sum rule.
(c) Relative  contributions of the three modes to the isovector sum rule.
The spin-isospin sound becomes the chiral soft mode.
}
\end{figure}

\newpage

\thispagestyle{empty}

\begin{figure}
\vspace{-12mm}
\epsfxsize = 9 cm
\centerline{\epsfbox{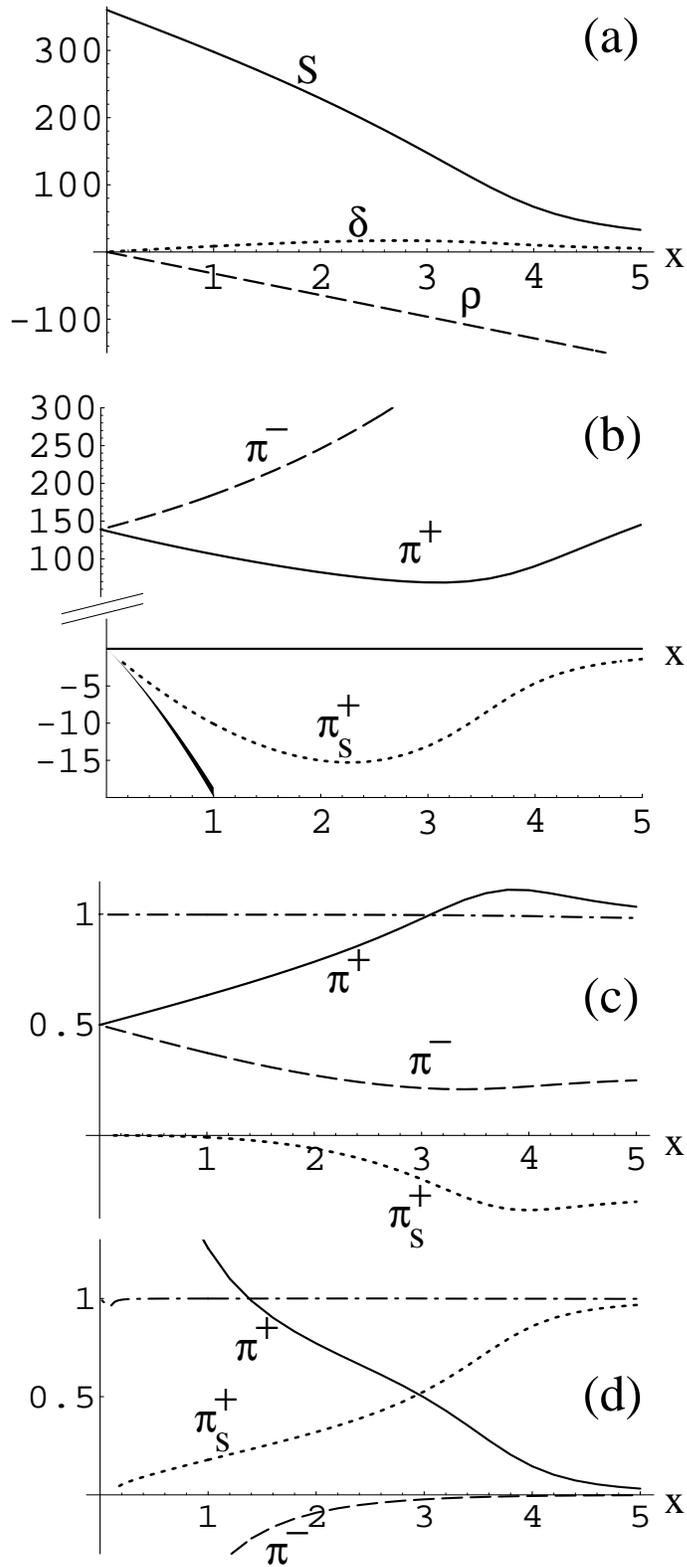}}
\vspace{-10mm}
\label{fig:setII}
\caption{Same as Fig.~1 for the parameter set II.
The spin-isospin sound mode in (b) has negative excitation energy.
}
\end{figure}

\newpage

\thispagestyle{empty}

\begin{figure}
\vspace{-12mm}
\epsfxsize = 9 cm
\centerline{\epsfbox{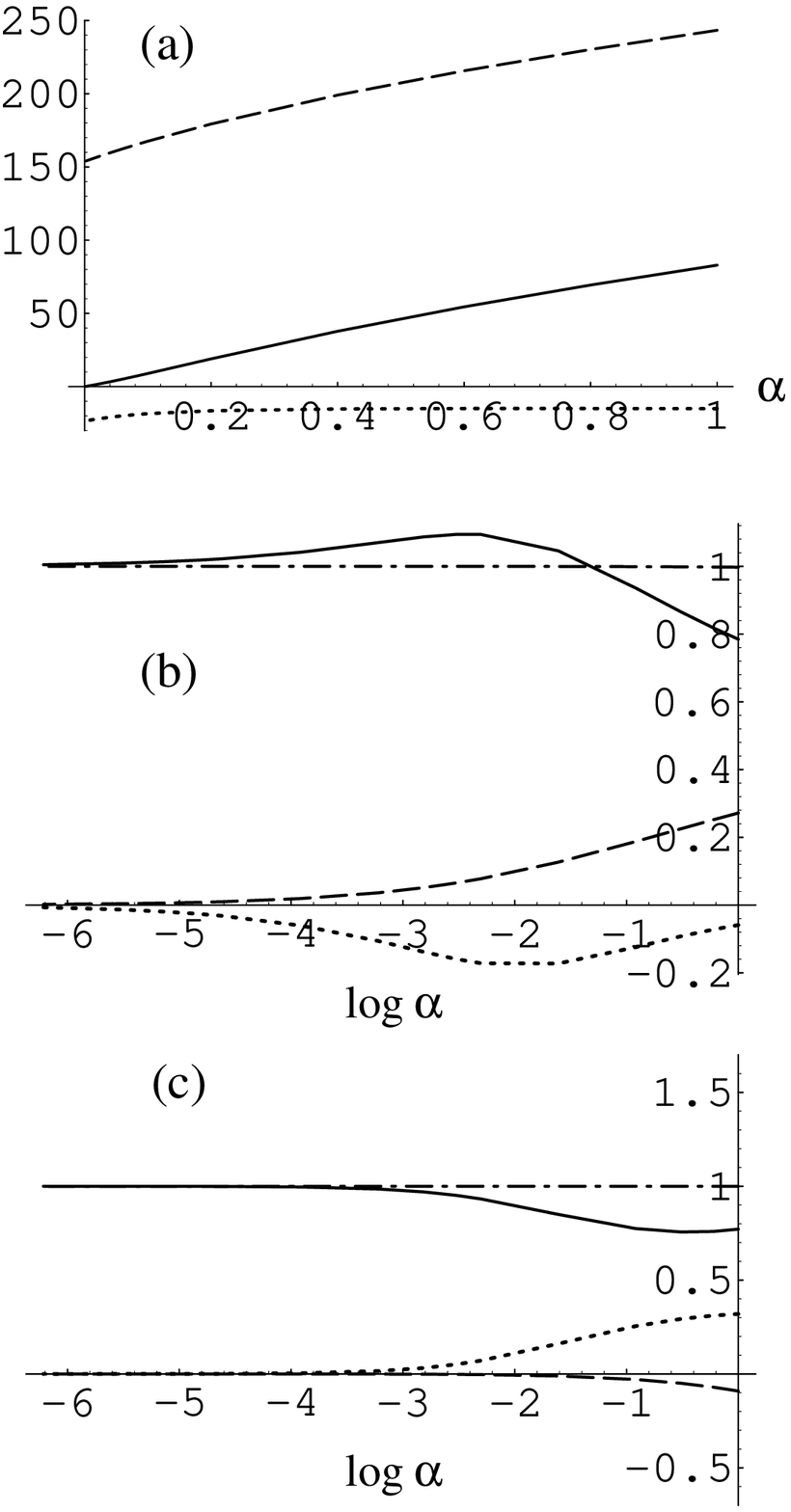}}
\vspace{-10mm}
\label{fig:chirII}
\caption{Same as Fig.~2 for the parameter set II.
The ``big'' mode becomes the chiral soft mode.
}
\end{figure}


\newpage

\begin{thebibliography}{9}

\bibitem{CB} T. D. Cohen and W. Broniowski, Phys. Lett. {\bf B 348}
(1995) 12.

\bibitem{Lutz}
M. Lutz, A. Steiner, and W. Weise, {Nucl. Phys.} {\bf A574} 
 (1994) 755.

\bibitem{NJLrev}
For recent reviews and references see:
T. Hatsuda and T. Kunihiro, Phys. Rep. {\bf 247} (1994) 221;
J. Bijnens, Phys. Rep. {\bf 265} (1996) 369.

\bibitem{migdal} 
A. B. Migdal, Rev. Mod. Phys. {\bf 50} (1978) 107, and references therein.

\bibitem{rhowil}
G. Baym and D. K. Campbell, in {\em Mesons and Nuclei}, p. 1031,
eds. M. Rho and D. H. Wilkinson (North Holland, 1979), and
references therein.

\bibitem{BernardAP}
V. Bernard, A. H. Blin, B. Hiller, Y. P. Ivanov, A. A. Osipov,
and U.-G. Meissner, Ann. Phys. (NY) {\bf 249} (1996) 499.

\bibitem{Klimt}
M. Lutz, S. Klimt, and W. Weise, {Nucl. Phys.} {\bf A542} 
 (1992) 521.

\bibitem{sao}
M. C. Ruivo and C. A. de Sousa, Phys. Lett. {\bf B 385} (1996) 39.

\bibitem{kubodera}
H. Yabu, S. Nakamura, F. Myhrer, and K. Kubodera,
Phys. Lett. {\bf B 315} (1993) 17.

\end{thebibliography}
\end{document}